\documentclass{aastex}
\usepackage{emulateapj5}
\usepackage{apjfonts}

\usepackage{graphics}

\newcommand{\bd}{\begin{displaymath}}
\newcommand{\ed}{\end{displaymath}}

\shorttitle{Constraints on radiatively inefficient accretion
history} \shortauthors{Cao}

\begin{document}


\title{Constraints on radiatively inefficient accretion
history of active galactic nuclei from hard cosmological X-ray
background}

\author{Xinwu Cao}
\affil{Shanghai Astronomical Observatory, Chinese Academy of
Sciences, 80 Nandan Road, Shanghai, 200030, China\\Email:
cxw@shao.ac.cn\\(Received 2005 July 13; accepted 2005 August 24)}

\begin{abstract}
The transition of a standard thin disk to a radiatively
inefficient accretion flow (RIAF) is expected to occur, when its
accretion rate $\dot{m}$ is lower than the critical value
$\dot{m}_{\rm crit}$ ($\dot{m}=\dot{M}/\dot{M}_{\rm Edd}$). The
RIAF is very hot, and it radiates mostly in the hard X-ray band
($\ga$ 100~keV). Assuming that the accretion disk in every bright
active galactic nucleus (AGN) will finally undergo a RIAF phase
while $\dot{m}<\dot{m}_{\rm crit}$, we calculate the contribution
of the RIAFs in AGNs to the cosmological X-ray background of
$10-1000$~keV. We find that the timescale $t^{\rm RIAF}$ of the
RIAF accreting at $\sim\dot{m}_{\rm crit}$ should be shorter than
$\sim 10^{-2}t^{\rm b}$ if $\dot{m}_{\rm crit}=0.01$, where
$t^{\rm b}$ is the lifetime of bright AGNs, i.e., $\dot{m}$
declines from $\dot{m}_{\rm crit}$ to a rate significantly lower
than $\dot{m}_{\rm crit}$ within $t^{\rm RIAF}$. The derived
timescale $t^{\rm RIAF}$ is affected by the parameters adopted in
the model calculations, which is also discussed in this {\it
Letter}.

\end{abstract}

\keywords{galaxies: active---quasars: general---accretion,
accretion disks---black hole physics; X-rays: diffuse background}


\section{Introduction}

The AGN X-ray luminosity function (XLF) is directly linked to the
accretion history of AGNs in the universe. Many works on XLFs were
carried out in either the soft X-ray band ($\la 3$~keV)
\citep*[e.g.,][]{m91,b93,p97,mhs00} or the hard X-ray band ($\ga
2$~keV) \citep*[e.g.,][]{b98,c03,u03}. The luminosity functions
(LFs) derived from the surveys in the soft X-ray band ($\la$
3~keV) may have missed many obscured (type II) AGNs, while the
hard X-ray surveys ($\sim$ 2-10~keV) can trace the whole AGN
population including obscured type II AGNs. The cosmological X-ray
background (XRB) is mostly contributed by AGNs \citep{h98,s98}. In
the most popular synthesis models of the XRB based on the
unification schemes for AGNs, the cosmological XRB from $\la
2$~keV to more than several hundred keV can be fairly well
reproduced by using a template spectrum of AGNs consisting of a
power-law X-ray spectrum with an exponential cutoff around several
hundred keV \citep*[e.g.,][]{mf94,mgf94,c95,grs99,u03}.
\citet{df97} alternatively proposed that the very hard X-ray
background (VHXRB, the term is used here for the hard X-ray
background above 10 keV to distinguish from the conventionally
mentioned hard X-ray background in 2$-$10 keV) may be dominated by
the thermal bremsstrahlung emission from the advection dominated
accretion flows (ADAFs) in low-luminosity AGNs. Further detailed
ADAF spectral calculations \citep{d99} showed that many sources at
redshift $z\sim 2-3$ with ADAFs accreting at the rates close to
the critical value are required to reproduce the observed VHXRB
spectral shape. Recently, \citet{u03} derived a hard X-ray
luminosity function (HXLF) from a highly complete AGN sample
(2-10~keV), which includes both type I and type II AGNs (except
Compton thick AGNs). Based on this HXLF, their synthesis models
can explain most of the observed XRB from the soft X-ray band to
the hard X-ray band around several hundred keV. Their calculations
slightly ($\approx$10--20\%) underestimate the relative shape of
the XRB spectrum around its peak intensity (see Fig. 18 in their
work). Such a discrepancy can be explained provided the same
number of Compton-thick AGNs with $\log N_{\rm H}=24-25$ as those
with $\log N_{\rm H}=23-24$ is included. This implies that the
contribution to the VHXRB from the radiatively inefficient
accretion flows (RIAFs) in AGNs may be important, but not
dominant, even if the contribution from the Compton-thick AGNs is
not considered.

There are a variety of studies exploring the evolution of AGNs
based on either optical quasar LFs or XLFs, or both
\citep*[e.g.,][]{hr93,hm00,kh00,yt02,wl03,ma04}. A common
conclusion is that the timescale of AGN activities is short
compared with the Hubble timescale, though the quantitative
results on the bright quasar lifetime vary from $\sim10^{7}$ to
$\sim10^{9}$ years  for different investigations.
The standard optically thick accretion disks are present in bright
AGNs, provided the accretion rate is high. The AGN activity may be
switched off while the gases near the black hole are exhausted
\citep*[see][for a recent review, and references therein]{n02}.
While the accretion rate $\dot{m}$ ($\dot{m}=\dot{M}/\dot{M}_{\rm
Edd}$) declines below a critical value $\dot{m}_{\rm crit}$, the
standard disk transits to a RIAF \citep*[e.g.,][]{ny95}. The RIAF
is optically thin, very hot, and its spectrum is peaked at around
several hundred keV. There are numerous observational evidences
indicating that the RIAFs are indeed present in many
low-luminosity AGNs and in our galactic center Sgr A$^*$
\citep*[e.g.,][]{ny95,l96,gnb99,yn04}.

In this {\it Letter}, we will explore how the inefficient
accretion history of AGNs is constrained by the VHXRB. The
cosmological parameters $\Omega_{\rm M}=0.3$,
$\Omega_{\Lambda}=0.7$, and $H_0=70~ {\rm km~s^{-1}~Mpc^{-1}}$
have been adopted in this work.

\section{Model}

The HXLF given by \citet{u03} is so far most suitable for our
present investigation, as it includes both type I and type II AGNs
(except Compton thick AGNs).  Assuming the accretion disk in every
bright AGN to transit to a RIAF while its accretion rate is low,
we derive the co-moving space number density of these faint AGNs
containing RIAFs from the HXLF, provided the accretion rate
evolution is known. Based on theoretical spectral calculations for
the RIAFs, strict constraints on the accretion history of AGNs can
be achieved from the comparison with the cosmological VHXRB in
$10-1000$~keV.

\subsection{Space density of faint AGNs}

We assume that all AGNs described by the HXLF have standard
accretion disks (we will come to justify this assumption in the
Discussion section). Hereafter, we refer to these AGNs with
standard disks as ``bright AGNs", while those AGNs with RIAFs as
``faint AGNs". The X-ray luminosity in $2-10$~keV can be converted
to the bolometric luminosity by using an empirical relation:
$L_{\rm bol}=f_{\rm cor}L_{\rm X}$, where $f_{\rm cor}=100$ is
adopted \citep{erz02,me04}. The black hole mass density for bright
AGNs in the co-moving space at redshift $z$ can be calculated by
\begin{equation} \rho^{\rm b}_{\rm bh}(z)={\frac {f_{\rm
cor}}{\dot{m}^{\rm aver}L_{\rm Edd,\odot}}}
\int\limits^{48}_{41.5}L_{\rm X}\frac{{\rm d} \Phi (L_{\rm X},
z)}{{\rm d Log} L_{\rm X}} {\rm d Log} L_{\rm X}~~~ {\rm
M}_{\odot} {\rm Mpc^{-3}}, \label{bhmdens}\end{equation} where
$\Phi (L_{\rm X}, z)$ is the HXLF given by \citet{u03},
$\dot{m}^{\rm aver}$ is the average dimensionless accretion rate
for bright AGNs described by this HXLF, and $L_{\rm
Edd,\odot}=1.38\times10^{38} {\rm ergs~s^{-1}}$ is the Eddington
luminosity for a black hole with solar mass.

In this {\it Letter}, we assume that the black hole mass does not
change significantly after the accretion mode transition, which is
satisfied only if we consider the time after the transition is
shorter than the Salpeter timescale, because the accretion rates
of these faint AGNs are very low ($\dot{m}\la10^{-2}$). The total
monochromatic X-ray luminosity of all faint AGNs in unit of
co-moving volume is given by \bd L_{\rm X}^{\rm f,tot}(z)=\int
n^{\rm f}(M_{\rm bh},z){\rm d}M_{\rm
bh}~~~~~~~~~~~~~~~~~~~~~~~~~~~~~~~~~~~~~~~~~~~~~~~~~~\ed
\begin{equation} \times \int\limits_{0}^{L_{\rm X}^{\rm
crit}(M_{\rm bh},E)}{\frac {1}{t^{\rm f}}}\left[{\frac {{\rm
d}L_{\rm X}(E,t)}{{\rm d}t}}\right]^{-1}L_{\rm X}(E){\rm d}L_{\rm
X}(E), \label{totlx}\end{equation} where $n^{\rm f}(M_{\rm bh},z)$
is the black hole mass function for faint AGNs, $L_{\rm X}^{\rm
crit}(M_{\rm bh},E)$ is the X-ray luminosity of the RIAF accreting
at $\dot{m}=\dot{m}_{\rm crit}$. For simplicity, we employ a
conventionally adopted assumption of a fixed lifetime $t^{\rm f}$
and the same light curve for all faint AGNs. For the RIAFs in
AGNs, their spectra $L_{\rm X}(E)$ depend almost linearly on the
black hole mass $M_{\rm bh}$, and Eq. (\ref{totlx}) can be
re-written as \bd L_{\rm X}^{\rm f,tot}(E,z)={\frac 1 {t^{\rm
f}}}\int {\frac {M_{\rm bh}}{10^{8}\rm M_\odot}}n^{\rm f}(M_{\rm
bh},z){\rm d}M_{\rm bh}\int\limits_{0}^{t^{\rm f}}l_{\rm
X}(E,t){\rm d}t\ed\begin{equation}={\frac 1 {t^{\rm f}}}{\frac
{\rho^{\rm f}_{\rm bh}(z)}{10^{8}\rm M_\odot}}
\int\limits_{0}^{t^{\rm f}}l_{\rm X}(E,t){\rm d}t,
\label{totlx2}\end{equation} where $l_{\rm X}(E,t)$ is the faint
AGN light curve for a $10^8{\rm M}_\odot$ black hole, and it can
be calculated provided the time-dependent accretion rate
$\dot{m}(t)$ is known. The total number density $N^{\rm f}(z)$ of
faint AGNs is: $N^{\rm f}(z)=N^{\rm b}(z)t^{\rm f}/t^{\rm b}$,
where $t^{\rm b}$ is the liftime of bright AGNs, and the bright
AGN number density $N^{\rm b}(z)$ can be calculated from the HXLF.
The black hole mass density $\rho_{\rm bh}(z)=N(z)M_{\rm bh}^{\rm
aver}$, where $M_{\rm bh}^{\rm aver}$ is the average black hole
mass. The average black mass for faint AGNs should be larger than
that for bright AGNs, as the black holes in bright AGNs are still
growing through accretion (a rough estimate can give the average
hole hole mass in faint AGNs being about twice of that in bright
AGNs, if $\dot{m}^{\rm aver}=1$ and $t^{\rm b}\sim 10^8$ years).
This leads to
\begin{equation} \rho^{\rm f}_{\rm bh}(z)={\frac {t^{\rm f}}{t^{\rm
b}}} \rho^{\rm b}_{\rm bh}(z)f_{\rm
bh},\label{relbhm}\end{equation} where $f_{\rm bh}>1$ is the ratio
of average black hole masses of faint AGNs to bright AGNs. The
black hole mass density for faint AGNs can be calculated from the
HXLF by using Eqs. (\ref{bhmdens}) and (\ref{relbhm}), so that Eq.
(\ref{totlx2}) can be re-written as
\begin{equation} L_{\rm X}^{\rm f,tot}(E,z)={\frac {\rho^{\rm b}_{\rm
bh}(z)}{10^{8}\rm M_\odot}} {\frac {f_{\rm bh}}{t^{\rm b}}}
\int\limits_{0}^{t^{\rm f}}l_{\rm X}(E,t){\rm d}t.
\label{totlx3}\end{equation}  Based on the RIAF models, the X-ray
light curve can be calculated, if we know how the accretion rate
$\dot{m}$ evolves with time. Unfortunately, we are still ignorant
of the detailed form of $\dot{m}(t)$. Here, we introduce a
timescale $t^{\rm RIAF}$ to describe the basic feature of the
evolution of RIAFs in these faint AGNs, \begin{equation}
\int\limits_{0}^{t^{\rm f}}l_{\rm X}(E,t){\rm d}t\simeq l_{\rm
X}(E,0)t^{\rm RIAF}=l_{\rm X}^{\rm crit}(E)t^{\rm RIAF}.
\label{riaf}\end{equation} This timescale $t^{\rm RIAF}$ describes
how fast the accretion rate of a RIAF declines from $\dot{m}_{\rm
crit}$ to a rate significantly lower than $\dot{m}_{\rm crit}$
after the accretion mode transition. Now, Eq. (\ref{totlx3})
becomes
\begin{equation} L_{\rm X}^{\rm f,tot}(E,z)={\frac {\rho^{\rm
b}_{\rm bh}(z)}{10^{8}\rm M_\odot}} {\frac {t^{\rm RIAF}f_{\rm
bh}}{t^{\rm b}}}l_{\rm X}^{\rm crit}(E).
\label{totlx4}\end{equation}

The contribution from the RIAFs in all faint AGNs to the
cosmological XRB can be calculated by \bd f_{\rm
X}(E)=\int\limits_{0}^{z_{\rm max}}  {\frac {L_{\rm X}^{\rm
f,tot}[(1+z)E,z]}{4\pi d_{\rm L}^2(1+z)}}{\frac {{\rm d}V}{{\rm
d}z}}{\rm d}z ~~~~~~~~~~~~~~~~~~~~~~~~~~~~~~~~~~~~~~~\ed
\begin{equation}~~~~~~~~={\frac 1{10^{8}{\rm M_\odot}}}{\frac {t^{\rm
RIAF}f_{\rm bh}}{t^{\rm b}}} \int\limits_{0}^{z_{\rm max}} {\frac
{\rho_{\rm bh}^{\rm b}(z)l_{\rm X}^{\rm crit}[(1+z)E]} {4\pi
d_{\rm L}^2(1+z)}}{\frac {{\rm d}V}{{\rm d}z}}{\rm d}z.
\label{xrb}
\end{equation}
In this {\it Letter}, we conservatively adopt $f_{\rm bh}=1$.

\subsection{RIAF spectra}

In order to calculate the contribution of RIAFs in faint AGNs to
XRB, we need to have the X-ray spectrum $l_{\rm X}^{\rm crit}(E)$
of a RIAF around a $10^8{\rm M}_\odot$ black hole accreting at the
critical rate $\dot{m}_{\rm crit}$. We employ the approach
suggested by \citet{m00} to calculate the global structure of an
accretion flow surrounding a Schwarzschild black hole (i.e.,
$a=0$) in general relativistic frame. All the radiation processes
are included in the calculations of the global accretion flow
structure \citep*[see][for the details and the references
therein]{m00}. Unlike \citet{m00}'s calculations, which are
limited to the cases without winds, we also calculate the cases
with winds. In the spectral calculations, the gravitational
redshift effect is considered, while the relativistic optics near
the black hole is neglected. This will not affect our final
results on the XRB, as the faint AGNs should have randomly
distributed orientations and the stacked spectra would not be
affected by the relativistic optics. \figurenum{1}
\centerline{\includegraphics[angle=0,width=10.0cm]{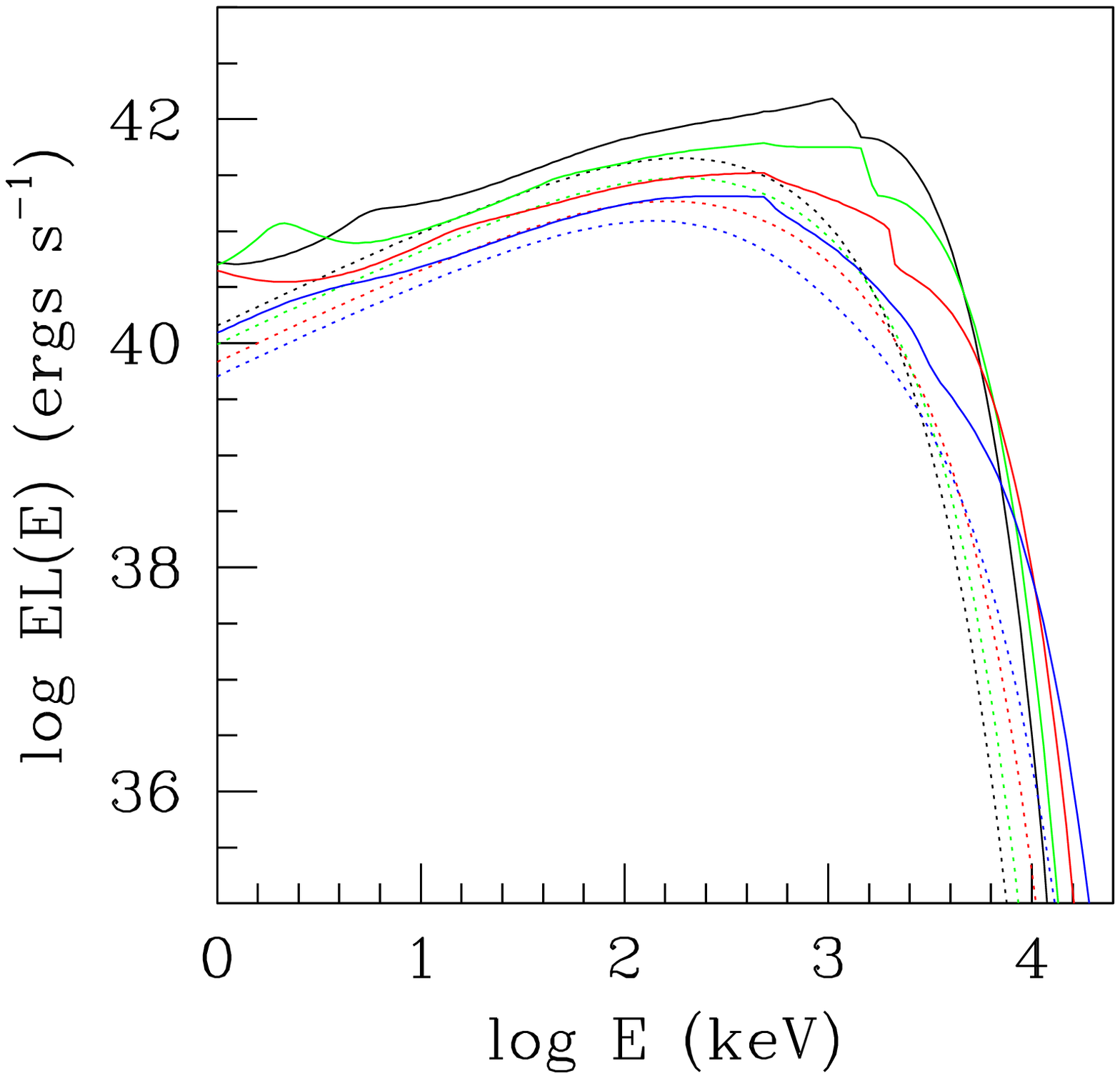}}
\figcaption{\footnotesize The spectra of RIAFs accreting at the
critical rate $\dot{m}_{\rm out}=0.01$, for different wind
parameters: $p_{\rm w}=0$(black), 0.2(green), 0.5(red), and
0.9(blue), respectively. The black hole mass $M_{\rm bh}=10^8{\rm
M}_\odot$, viscosity parameter $\alpha=0.2$, the fraction of the
magnetic pressure $1-\beta=0.5$, and the outer radius $R_{\rm
out}=100R_{\rm Schw}$ are adopted in the calculations. The dotted
lines represent the bremsstrahlung spectra of the RIAFs.
\label{fig1}} \centerline{}

\section{Results}

The three-dimensional MHD simulations suggest that the viscosity
parameter $\alpha$ in the accretion flows is $\sim 0.1$
\citep{a98}, or $\sim 0.05-0.2$ \citep{hb02}. We assume a
$r$-dependent accretion rate $\dot{m}=\dot{m}_{\rm out}(r/r_{\rm
out})^{p_{\rm w}}$ for the RIAFs with winds. The parameters
adopted in the calculations are: $\alpha=0.2$, $\dot{m}_{\rm
out}=0.01$, the fraction of magnetic pressure $1-\beta=0.5$
($\beta=p_{\rm gas}/p_{\rm tot}$), and the fraction of dissipated
energy directly heating electrons is: $\delta=0.5$. The outer
radius $r_{\rm out}=100R_{\rm Schw}$ is adopted in all our
calculations, where $R_{\rm Schw}=2GM_{\rm bh}/c^2$. We plot the
X-ray spectra for RIAFs in Fig. \ref{fig1}.

There is no doubt that the contribution from a normal bright AGN
is important in the VHXRB, as the {\it BeppoSAX} observations
showed that the power-law X-ray spectra of bright AGNs extends to
several hundred keV \citep[e.g.,][]{n00}. Here we consider that
the XRB consists not only of the emission from type I/II bright
AGNs (Compton-thin) described by the HXLF, but also of the
emission from RIAFs in those faint AGNs which are not included in
the HXLF. In Fig. \ref{fig2}, we plot the results for the RIAFs
with different wind parameters $p_{\rm w}$ for different ratios
$t^{\rm RIAF}/t^{\rm b}$ ($\dot{m}^{\rm aver}=0.1$ is adopted for
bright AGNs). \figurenum{2}
\centerline{\includegraphics[angle=0,width=10.0cm]{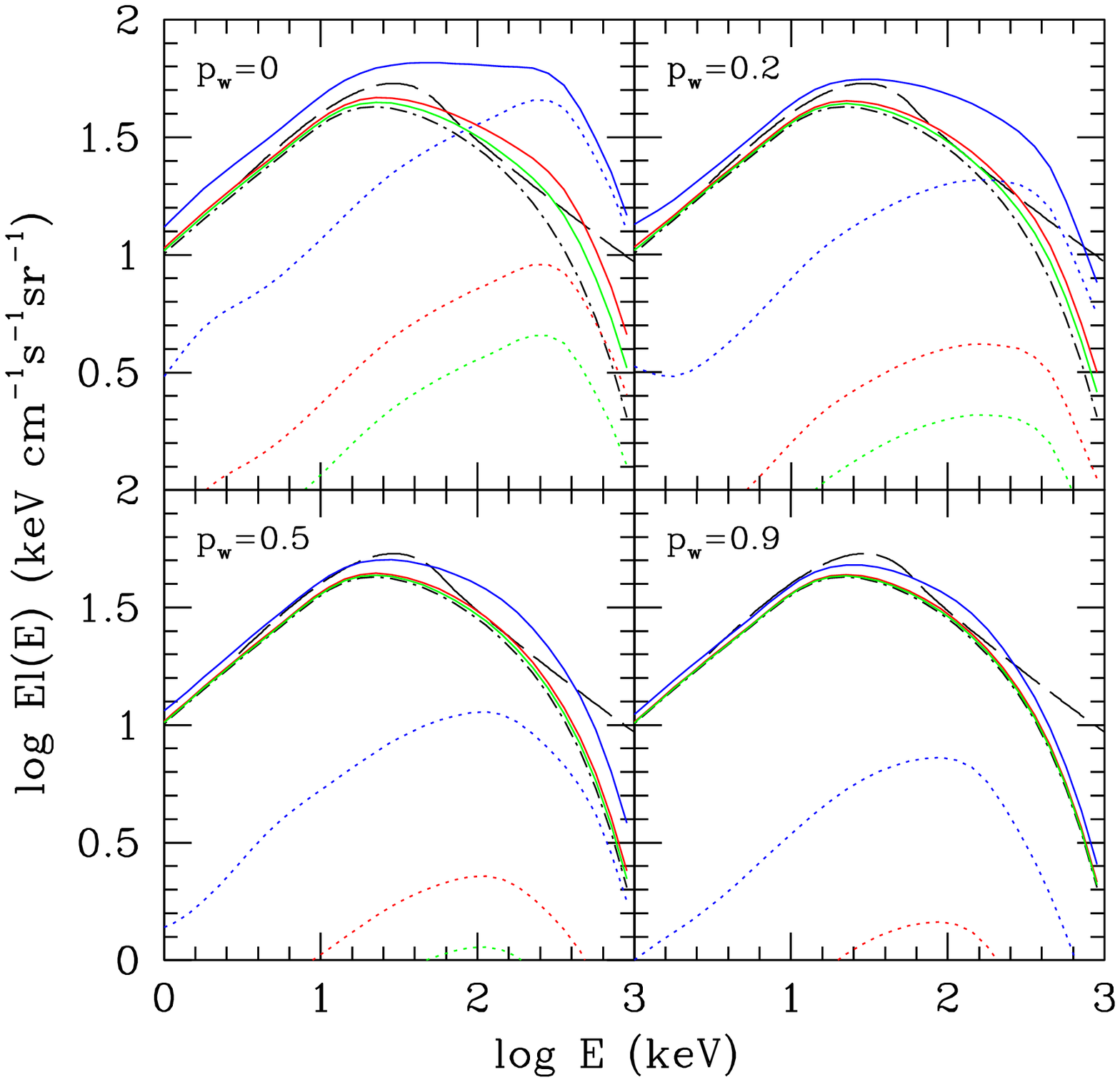}}
\figcaption{\footnotesize The contribution to XRB from bright and
faint AGNs for the RIAFs with different wind strengths or without
winds. The dashed lines are the observed XRB. The lines with
different colors represent different values of $t^{\rm
RIAF}/t^{\rm b}$=0.005(green), 0.01(red), and 0.05(blue),
respectively. The dot-dashed line represents the contribution by
bright type I/II AGNs (Compton-thin), which is taken from
\citet{u03}. The solid lines represent the XRB contributed by
bright type I/II (Compton-thin) AGNs and the RIAFs in the faint
AGNs, while the dotted lines are for the contributions of RIAFs in
the faint AGNs only.  \label{fig2}} \centerline{}

\section{Discussion}

We find that all spectra have an energy peak at around  several
hundred to 1000~keV (see Fig. \ref{fig1}). Compared with those
obtained by \citet{d99}, our spectra have higher energy peaks. The
reason is that we adopt $\delta=0.5$, larger than theirs, which
leads to higher electron temperatures of the accretion flows. From
Fig. \ref{fig2}, we find that $t^{\rm RIAF}\la 0.05t^{\rm b}$ is
required from the comparisons of our theoretical calculations with
the observed XRB, for any RIAFs either without winds or with
strong winds. For the RIAFs without winds ($p_{\rm w}=0$) or weak
winds ($p_{\rm w}=0.2$), we find that the hard X-ray emission
alone from the RIAFs in faint AGNs have already surpassed the
observed XRB, provided $t^{\rm RIAF}=0.05t^{\rm b}$. If the hard
X-ray emissions from both the bright type I/II AGNs and RIAFs in
faint AGNs are considered, we find more strict constraints on the
RIAF timescale: $t^{\rm RIAF}\la 0.01t^{\rm b}$ required for most
cases (it becomes $t^{\rm RIAF}\la 0.005t^{\rm b}$ for the RIAFs
without winds). Here, we have neglected the contribution from the
Compton-thick AGNs. If the contribution from a Compton-thick AGN
is included, the present derived RIAF accretion timescales will
become even lower.

Our present calculations are based on the assumption that all AGNs
described by the Ueda's HXLF are bright, i.e., the standard bright
accretion disks are responsible for their energy sources. Our RIAF
spectral calculations show that $L_{\rm X}^{2-10\rm
keV}=1.48\times 10^{41}\rm ergs~ s^{-1}$ for $p_{\rm w}=0$, and
$5.40\times 10^{40}\rm ergs~ s^{-1}$ for $p_{\rm w}=0.9$,
respectively, if $M_{\rm bh}=10^8{\rm M}_\odot$ and $\dot{m}_{\rm
out}=0.01$. As the lower luminosity limit of the HXLF extends to
$10^{41.5}$\rm ergs s$^{-1}$, it means that only the faint AGNs
accreting at the critical rate with black hole masses $\ga
2\times10^8{\rm M}_\odot$ may have appeared in this HXLF. The
standard disks accreting at $\dot{m}>\dot{m}_{\rm crit}$ around a
black hole with $\ga 2\times10^8{\rm M}_\odot$ have $L_{\rm
X}^{\rm 2-10keV}\ga 10^{42.5}$ ergs~s$^{-1}$.  This implies that
some RIAF counterparts of the bright AGNs with $L_{\rm X}^{\rm
2-10keV}\ga 10^{42.5}$ ergs~s$^{-1}$ may be included in this HXLF.
We can roughly estimate that the number ratio of these RIAF
counterparts to bright AGNs with $L_{\rm X}^{\rm 2-10keV}\ga
10^{42.5}$ ergs~s$^{-1}$ is: $\la t^{\rm RIAF}/t^{\rm b}$. Simply
integrating the HXLF, we find that the ratio of the sources with
$L_{\rm X}^{\rm 2-10keV}=10^{42.5-48}\rm ergs~s^{-1}$ to all AGNs
described by this HXLF is about 0.13, which implies that less than
a fraction $\sim 0.13t^{\rm RIAF}/t^{\rm b}$ of all sources with
$L_{\rm X}^{\rm 2-10keV}=10^{41.5-48}\rm ergs~s^{-1}$ may have
RIAFs. Therefore, the contribution to the XRB from those RIAF
sources with $L_{\rm X}^{\rm 2-10keV}>10^{41.5}\rm ergs~s^{-1}$
may be over-calculated, but at a very low level of $\la0.13t^{\rm
RIAF}/t^{\rm b}$, which is negligible and will not affect our main
conclusions.

We have assumed a constant lifetime $t^{\rm b}$ and $t^{\rm f}$
($t^{\rm RIAF}$) for all bright and faint AGNs, which may not be
true, as suggested by \citet{h05c}. However, our results on the
XRB only depends on the ratio $t^{\rm RIAF}/t^{\rm b}$, so our
conclusions will not be altered, even if each individual sources
have different lifetimes, provided they have a similar
time-dependent form of $\dot{m}(t)$. The resulted XRB from Eq.
(\ref{xrb}) depends on the value of $\dot{m}^{\rm aver}$ as
$\propto 1/\dot{m}^{\rm aver}$ (see Eq. \ref{bhmdens}).
\citet{md04} estimated that the average accretion rate
$\dot{m}^{\rm aver}$ varies from 0.1 at $z\sim0.2$ to 0.4 at
$z\sim2$ from a large sample of SDSS quasars. If we adopt a larger
$\dot{m}^{\rm aver}=0.4$, the derived RIAF timescales $t^{\rm
RIAF}$ will be four times of the present values. The derived
timescales are proportional to the bolometric luminosity
correction factor $f_{\rm cor}$, and the uncertainty of $f_{\rm
cor}$ should not be very large, which will not affect our main
conclusions. The emissions in hard X-ray bands from RIAFs are
dominated by the bremsstrahlung emission and Comptonization of the
bremsstrahlung photons (see Fig. \ref{fig1}), which are almost
independent of $\beta$ (magnetic field strength) adopted in the
calculations.

Our results indicate that the accretion rate must drop to a very
low rate from the critical accretion rate in a short timescale
compared with the bright AGN lifetime. Recent hydrodynamical
simulations on the evolution for the AGN formed after a merger
indeed show a rapid decline of accretion rate from near Eddington
rate to $\sim 10^{-5}$ within a short timescale compared with its
bright phase \citep*[see Fig. 2 in][]{d05}, which is qualitatively
consistent with our results. The black hole growth during the
timescale $t^{\rm RIAF}$ can be neglected compared with that
during the bright AGN phase, as $t_{\rm RIAF}\ll t^{\rm b}$.
However, our present calculations have not considered how the
accretion rate evolves with time in detail. The advection becomes
important while the RIAFs are accreting at rates $\ll\dot{m}_{\rm
crit}$, the black holes may swallow gases without radiating much
X-rays. So, it cannot be ruled out the possibility that the faint
AGNs stay at accretion rates far lower than the critical value for
a very long time, say, comparable with the Hubble timescale. In
principle, this can also be constrained by the cosmological XRB,
which is beyond the scope of this {\it Letter}.

\acknowledgments  I am grateful to R. Narayan for stimulating
discussion, insightful comments and suggestions. I thank the CfA
for its hospitality, where this work was finally carried out. This
work is supported by the National Science Fund for Distinguished
Young Scholars (grant 10325314), the NSFC (grant 10333020), and
the NKBRSF (grant G1999075403).

\end{document}